\documentclass{openjournal}

\usepackage{amsmath, verbatim}
\usepackage[multidot]{grffile}
\usepackage{lineno}

\usepackage[hyperindex,breaklinks]{hyperref}
\usepackage[anythingbreaks]{breakurl}

\begin{document}

\title{Characterizing the Sample Selection for Supernova Cosmology}
\author{Alex G.\ Kim, for the LSST Dark Energy Science Collaboration}
\affiliation{Physics Division, Lawrence Berkeley National Laboratory, Berkeley, CA 94720, USA}
\begin{abstract}
Type Ia supernovae (SNe Ia) are used as distance indicators to infer the cosmological parameters that specify the expansion history of the universe.
Parameter inference depends on the criteria by which the analysis SN sample is selected.
Only for the simplest  selection criteria and population models can the likelihood be calculated
analytically, otherwise  it needs to be determined numerically, a process that inherently has error.
Numerical errors in the likelihood lead to errors in parameter
inference.  
This article presents  toy examples where the distance modulus is inferred given a set of SNe at a single redshift.
Parameter estimators and their uncertainties are calculated using Monte Carlo techniques.
The relationship between the number of Monte Carlo realizations and numerical errors is presented.
The procedure can be applied to more realistic models and used to determine the computational and data management
requirements of  the  transient analysis pipeline.

\end{abstract}
\makeatletter
\let\start@align@nopar\start@align
\let\start@gather@nopar\start@gather
\let\start@multline@nopar\start@multline
\long\def\start@align{\par\start@align@nopar}
\long\def\start@gather{\par\start@gather@nopar}
\long\def\start@multline{\par\start@multline@nopar}
\makeatother
\section{Introduction}
In the early days of supernova cosmology, hard redshift cuts were applied to
spectroscopically classified Type~Ia supernovae (SNe~Ia) in order   to mitigate against Malmquist bias and biases due to non-Ia contamination \citep{1998AJ....116.1009R,1999ApJ...517..565P}.
Malmquist bias arises because SN~Ia peak magnitudes have an intrinsic dispersion; if  in a magnitude-limited survey intrinsically faint undetected SNe are
unaccounted for, inferred distances can be underestimated.
Most other  transients discovered in imaging surveys are intrinsically fainter than SNe~Ia, and
including these objects in an analysis as if they were SNe~Ia would  lead to  overestimated inferred distances.
In the past,
strict sample-selection cuts made these sources of bias negligible, though at the expense of reducing the sample size.

A motivation for relaxing the sample-selection criteria is to extend the redshift range of the SN~Ia Hubble diagram
and to take advantage of the large number of SN discoveries from modern supernova surveys.  
SNe at higher redshift have fainter observed magnitude, making them more susceptible  to Malmquist bias and
more difficult to classify spectroscopically; given finite follow-up resources we will have to 
rely on photometric classification  \citep{2013ApJ...763...88C,2018ApJ...857...51J} in order to consider all discoveries. 
Analysis models then need to accommodate not only SNe~Ia but also a contaminant population
\citep{2007PhRvD..75j3508K}.
Under these circumstances, the effect of sample selection
cannot be ignored \citep{2015ApJ...813..137R,2017arXiv170603856H,2019ApJ...876...15H}.

The criteria for sample selection are becoming more sophisticated.  Certain criteria, such as those based on signal-to-noise,
color, and data quality,  are simply expressed as a function of the observed data.
Machine learning classifiers are being developed
to select likely SNe~Ia \citep{2012MNRAS.419.1121R, 2016ApJS..225...31L, 2017ApJ...837L..28C, 2018MNRAS.473.3969R, 2019ApJ...884...83V, 2019AJ....158..257B, 2019PASP..131k8002M, 2019A&A...627A..21P,
2020JAI.....950005G, 2020MNRAS.491.4277M}, but their inner workings  can be opaque and not easily modeled
as a simple function of data.
Determining whether a supernova enters the sample requires nothing short of running its data through the
black box classifier.
Without an analytic expression for the sample selection, the volume of classifications required to characterize
the classifier to achieve a targeted precision in parameter inference could be computationally non-trivial.

An added complication in modern surveys is that  sample selection may not entirely defined by the
cosmology analysis team.  For example, the transients discovered by the
Vera C.\ Rubin Legacy Survey of Space and Time (LSST) are expected to undergo a series of pipelines, each
the separate responsibility of either the Project,
Broker teams, or  Science Collaborations. 
The LSST Project plans to identify transients in cadenced imaging surveys, triggering
alerts that will be distributed to a small number of third-party brokers who
will distribute filtered subsets to the community in near-real-time.  In addition, all
alerts and objects will be available within 24 hours of shutter close through the Project's Prompt Data Products databases.\footnote{
Relevant LSST documents are available at \url{https://lse-163.lsst.io/}, \url{https://ldm-612.lsst.io/}.}
Science teams that use these products to define their samples may find that they need to understand the
sample-selection efficiencies of antecedent processes that are the responsibility of other parties.

Supernova cosmology analysis for a contaminated sample has been described in the literature.
Hierarchical models for the supernova populations, observations, and the sample selection are used to derive the 
likelihoods needed for parameter inference \citep{2015ApJ...813..137R, 2019ApJ...876...15H}.  
There are a number of analyses in progress using this approach, though they have yet to be published.
BEAMS and its extensions
\citep{2012ApJ...752...79H, 2013JCAP...01..039K, 2017JCAP...10..036R} present a likelihood given the known probabilities of type.
However, this
likelihood is not necessarily applicable 
in practice as only an estimator, and not the true probabilities, are known, which for efficient classifiers 
that use brightness and color information are
covariant with the inferred distances.
In \citet{2012ApJ...752...79H} the impact of using incorrect probabilities is explored, but such  tests on simulated data
do not translate to validation of real data.
The models underlying the simulation may not represent the astrophysics of the real Universe to high fidelity.
This  problem is not unique
to BEAMS and affects all model inference.  The difficulty is that proper interpretation of BEAMS cosmology results depends on understanding
a simulation whose inner tunings are subtle and not necessarily known by the bulk of the community.
The cosmology analyses of PanSTARRS and DES \citep{2018ApJ...857...51J, 2019ApJ...874..150B},
based on ``BEAMS with Bias Corrections'' \citep{2017ApJ...836...56K}, 
add systematic effects and bias corrections in an ad hoc manner, not from first principles.
The resulting ``likelihood'' is not a likelihood, as it does not include the probabilities of detection.
This is readily apparent in that the probability distribution function (pdf) of observables of an experiment with sample selection is in general not normal,
whereas  their pdf's are normal; the implication is that the confidence regions are not based on a likelihood.
Note  that the confidence-region test in  \citep{2019ApJ...874..150B} quantifies the range of data possible
from one point in parameter space, not the range of parameter space consistent with a single data realization upon which frequentist
confidence intervals are based.

Apparent in the hierarchical models is the sensitivity of the likelihood to the physics model and the choices
made for sample selection.  By the time LSST SN discoveries are used for cosmology analysis, SNe of all types should
be better understood and modeled  than they are today.
In addition, the algorithms
 that will be used for triggered follow-up and cosmology sample selection are currently being evaluated.  Indeed, a motivation for the current work is to identify
important metrics that describe the computing load engendered by different  sample selection pipelines.
For these reasons, this article focuses on the
methodology and for clarity uses simple models as examples.  Implementing this methodology for use on a suite of models
and sample-selection (e.g., classifier) pipelines is  planned.

This article  presents a procedure to determine errors in parameter inference due to numerical errors that  naturally arise
when taking into account sample selection.
Up until now, the  sample selection in SN~Ia cosmology analysis
could be effectively characterized by the properties of the measurement from a single observation or visit (e.g.,
signal-to-noise, percent flux increase, magnitude) or be otherwise
due to processes independent of the source properties (e.g., mis-subtractions,
lack of spectroscopy follow-up time).  Now that new surveys discover large numbers of non-SN~Ia transients,
the sophisticated classifiers now included as part of analysis pipelines complicate sample-selection characterization.
The motivation of this work is to determine how often classifiers have to be queried in order  to achieve precise distance inference.
Though we are motivated by LSST, the procedure is of general applicability.
In \S\ref{model:sec} we present  a generic model, its likelihood, and the
best-fit parameters and  uncertainties calculated  from the likelihood, for a survey with sample selection.
Monte Carlo integration is introduced to calculate the numerical integrals involved.
For toy examples, in \S\ref{examples:sec}
we propagate  numerical-integration errors into those of   the best-fit parameters and parameter uncertainties, as a function of the number of Monte Carlo
samples.   A discussion on those results and their implications on  computing resources are given in  \S\ref{discussion:sec}.
Conclusions are given in \S\ref{conclusions:sec}.

\section{Distances from a contaminated sample of standard candles}
\label{model:sec}
\subsection{The model}
This article considers a simple model that captures the main complications of future SN~cosmology analyses.
We consider standard candles each with a measured distance estimate, without reference to the processes of getting that estimate
from observables. 
All the objects have the same distance, which is the limit of considering a subset of SNe that lie within a narrow redshift bin.
There are two populations of objects, one brighter candle with tight intrinsic magnitude dispersion and a second fainter ``candle''  with
a broad magnitude dispersion. 
The observed sample of objects is magnitude limited, replicating a detection threshold in their discovery.
The discovered sample undergoes further classification filtering, using supplemental data and potentially the distance measurement, in an effort
to use only objects from the bright population.  The classifier may not be perfect, filtering out some of the desired population while leaking
in some of the undesired fainter population.
The model has two top-level parameters: the distance of the objects and the underlying ratio in the numbers of the two populations.

Here we summarize the assumptions, parameters, parameter dependencies, and notation for the model.
\begin{itemize}
\item All objects are at the same distance denoted by their distance modulus $\mu$.
\item There are two populations labeled as $T=0$, 1.  The $T=0$ population is of primary interest, being bright and precise distance
indicators 
 (e.g., SNe~Ia)
while the $T=1$ population represents the background (e.g., core-collapse supernovae).
The underlying fraction of type $T=0$ is given by $p_0$, such that
$P(T|p_0)= p_0 \delta_{T0}+(1-p_0)\delta_{T1}$, where $\delta$ is the Kronecker $\delta$.
\item All objects have a magnitude observable  $m$. The magnitude datum from a candle of population $T$ is drawn from a Normal distribution
\begin{equation}
p(m| T, \mu) =   \mathcal{N}(m-\mu-M_T,\sigma_T),
\end{equation}
where $\mu$ is the distance modulus, $M_T$ the candle's absolute magnitude, and $\sigma_T$ the  standard deviation, which is due to a combination
of intrinsic
dispersion and measurement noise.
Objects may also have additional data $x$ that are not sensitive to $\mu$ but are used in object detection
and classification, e.g., colors, light-curve shape, spectra, host-galaxy properties.

The above model for perfect standard candles does not apply to SNe~Ia, which are considered to be a family of similar objects
where the absolute magnitude $M_T$ of each individual SN is a parameter that can be inferred from data.
The intrinsic magnitude dispersion of SNe is not necessarily Gaussian.  In addition, foreground
effects e.g., dust extinction and, gravitational lensing, also affect observed magnitudes.
\item The data from different objects are independent.
\item 
Candles entering the sample must exceed a brightness threshold $m_\text{lim}$.  Each candle has a parameter $S=1$, 0
indicating whether it does or does not meet the brightness criterion. $P(S | m) = \delta_{S S(m)}$ where 
$S(m)= \mathcal{H}(m_\text{lim}-m)$  and $\mathcal{H}$ is the Heaviside function. 
In terms of LSST, this is akin to saying that the Project will efficiently generate magnitude-limited transient alerts.
\item 
Candles entering the sample are typically classified as having a high probability of being $T=0$.
A classifier ingests data and outputs results used to decide whether to include an object in the final sample.  The classification decision
is given by
$\tau(m, x)=1$, 0
for an object included/excluded from the sample.  $P(\tau | m, x) = \delta_{\tau \tau(m, x)}$.  For LSST the classification process
may be composed of several distinct components including live Public Broker assessments, real-time observing decisions, and a posteriori
classifications based on the full light curve.

The performance of the classifier  is described by its efficiency $\epsilon_0$ that a candle of  $T=0$ has $\tau=1$,
and its false-positive probability $\epsilon_1$ that a  candle of $T=1$ has $\tau=1$.
The probability of object of type $T$ being classified as $\tau$ is
$P(\tau|T)= (\epsilon_0 \delta_{0T} + \epsilon_1 \delta_{1T})\delta_{\tau 1}
+ (1-\epsilon_0 \delta_{0T} - \epsilon_1 \delta_{1T})\delta_{\tau 0}$.

In place of a real classifier, in  the  examples presented in this article the classifier efficiency and false-positive probability are taken to be random, hence completely uncorrelated
with $m$.
In this case the classifier itself does not  induce a bias in distance modulus.  Most classifiers, however, are expected
to use magnitude information and the presented methodology corrects for any potential bias.
\end{itemize}

\subsection{Parameter estimators and uncertainties}

In this work, the likelihood and its calculation are split into two more basic building blocks.
The first is  the expected fraction of all objects that pass both detection and classification criteria to make it into the analysis sample
\begin{align}  
\bar{S}(\mu, p_0)
& =p( S =1,\tau=1 |  \mu, p_0).
\end{align}
As will be shown shortly, it is the calculation of this term and its partial derivatives with respect to the parameters
that can require a large number of evaluations of the sample selection function to consider all objects that could have
entered the sample.
The second is the probability of an object with magnitude $m$ making it into the
analysis sample.
\begin{align}  
R(m,\mu, p_0)
& = p(m, S=1, \tau=1|  \mu, p_0).
\end{align}
This term requires calculating the sample selection probability of only those objects that are in the sample.

The likelihood is the probability given by the model for the sample-selected data, which is predicated on the detection and classification selection criteria being satisfied.
Recalling the independence of the data across objects
\begin{align}  
L(\mu, p_0; \{m\}) & = \prod_{i=1}^N p(m_i | S_i=1,\tau_i=1,  \mu, p_0) \\
& = \prod_{i=1}^N \frac{
p(m_i , S_{i}=1,{\tau_i}=1|  \mu, p_0)  }
{ p( S_{i}=1,{\tau_i}=1 |  \mu, p_0) } \\
& = \bar{S}(\mu, p_0) ^{-N} \prod_{i=1}^N 
R(m_i, \mu, p_0)  .
\end{align}
where the index $i$ runs over the $N$ objects in the final sample.
All objects contribute a common term for the probability of an arbitrary object making into the sample $\bar{S}$
and an individual term of its own probability of discovery $R$.

The estimators for parameters $\hat{\theta}$, $\theta \in \{\mu, p_0\}$, solve $\frac{\partial \ln{L}} {\partial \theta} =0$, i.e.\ 
\begin{align}  0=
- \frac{1}{\bar{S}} \frac{\partial \bar{S}}{\partial \theta}
+ \frac{1}{N} \sum_{i=1}^N \frac{1}{R(m_i)} \frac{\partial R(m_i)}{\partial \theta}.
\label{estimator:eqn}
\end{align}
The Hessian of the likelihood surface is
\begin{align}
H_{ij} &  = - \frac{\partial^2 \ln{L}} {\partial \theta_i \partial \theta_j} \nonumber \\
& = - \sum_{k=1}^N \left( \frac{1}{R(m_k)} \frac{\partial^2 R(m_k)}{\partial \theta_i \partial \theta_j} -  \frac{1}{R^2(m_k)} \frac{\partial R(m_k)}{\partial \theta_i} \frac{\partial R(m_k)}{\partial \theta_j} \right)  \nonumber \\
& + N \left( \frac{1}{\bar{S}} \frac{\partial^2 \bar{S}}{\partial \theta_i \partial \theta_j} - \frac{1}{\bar{S}^2} \frac{\partial \bar{S}}{\partial \theta_i} \frac{\partial \bar{S}}{\partial \theta_j} \right).
\label{hessian:eqn}
\end{align}
Its value at the extremum provides an estimate of the parameter uncertainty. In this article, the statistical uncertainty
in $\mu$ is given by $\sigma_\mu \approx \sqrt{H^{-1}_{\mu \mu}}$.
The estimators and their uncertainties are subject to error when there are uncertainties in the
logarithmic derivatives of $\bar{S}$.
The estimator depends on the average value of $\partial \ln{R} / \partial \theta$,  but is otherwise independent of $N$.
The Hessian similarly scales with $N$.  The errors in the parameter estimators
and the fractional errors in the parameter uncertainties presented in this article are thus independent of the sample size.

Given the model parameters and their probabilities, the  functions $\bar{S}$, $R$, and their partials can be expressed including the latent variables $T$ and $m$ as
\begin{align}  
\bar{S}(\mu, p_0)
& =\int   \sum_{T=0}^1 p( S=1,{\tau}=1, m, T |  \mu, p_0)   dm \nonumber \\
\frac{\partial \bar{S}}{\partial \mu}
& =\int  \sum_{T=0}^1 \frac{m-M_T-\mu}{\sigma_T^2}p( S=1,{\tau}=1, m, T |  \mu, p_0)   dm \nonumber \\
\frac{\partial \bar{S}}{\partial p_0}
& =\int  \sum_{T=0}^1 \frac{\delta_{T0}- \delta_{T1}}{p_0 \delta_{T0}+(1-p_0)\delta_{T1}}p( S=1,{\tau}=1, m, T |  \mu, p_0)   dm\nonumber  \\
\frac{\partial^2 \bar{S}}{\partial \mu^2}
& =\int   \sum_{T=0}^1 \left( \left( \frac{m-M_T-\mu}{\sigma_T^2}\right)^2 -\frac{1}{\sigma_T^2} \right)p( S=1,{\tau}=1, m, T |  \mu, p_0)   dm \nonumber \\
\frac{\partial^2 \bar{S}}{\partial \mu \partial p_0}
& =\int   \sum_{T=0}^1 \left( \frac{m-M_T-\mu}{\sigma_T^2} \right) \left( \frac{\delta_{T0}- \delta_{T1}}{p_0 \delta_{T0}+(1-p_0)\delta_{T1}} \right)p( S=1,{\tau}=1, m,  T |  \mu, p_0)   dm
\end{align}
and
\begin{align}
R(m, \mu, p_0)
& =  \sum_{T=0}^1   p(m, S=1,{\tau}=1, T|  \mu, p_0) \nonumber \\
\frac{\partial R }{\partial \mu}
& =\sum_{T=0}^1    \frac{m-M_T-\mu}{\sigma_T^2} p(m, S=1,{\tau}=1, T|  \mu, p_0) \nonumber \\
\frac{\partial R }{\partial p_0}
& =\sum_{T=0}^1    \frac{\delta_{T0}- \delta_{T1}}{p_0 \delta_{T0}+(1-p_0)\delta_{T1}}   p(m, S=1,{\tau}=1, T |  \mu, p_0) \nonumber \\
\frac{\partial^2 R }{\partial \mu^2}
& =\sum_{T=0}^1    \left( \left( \frac{m-M_T-\mu}{\sigma_T^2}\right)^2 -\frac{1}{\sigma_T^2} \right) p(m, S=1,{\tau}=1, T|  \mu, p_0) \nonumber \\
\frac{\partial^2 R }{\partial \mu \partial p_0}
& =\sum_{T=0}^1   \left( \frac{m-M_T-\mu}{\sigma_T^2} \right) \left( \frac{\delta_{T0}- \delta_{T1}}{p_0 \delta_{T0}+(1-p_0)\delta_{T1}} \right)  p(m, S=1,{\tau}=1,T |  \mu, p_0).
\end{align}
The second derivatives with respect to $p_0$ are  zero.
Although not denoted explicitly, the supplement data $x$  used only for detection and classification enter $\bar{S}$ and $R$ implicitly through $P(\tau | m,x)$.

\subsection{Monte Carlo integration and its errors}

$\bar{S}$ and its derivatives involve an integral whose integrands depend on the classifier through $\tau( m,x)$ and the detection selection
$S(m,x)$.  In general these functions
are challenging to express analytically with  precision.  Even if the integrands were analytic, the integrals themselves
are generally non-analytic.  The integrals are solved numerically through Monte Carlo integration, with the integrand evaluated
at randomly drawn sets of $\{m,x\}$ values.  The precision of the integration is 
dependent on the number of times the integrand is sampled.  Therein lies the technical motivation of this work, the requirements on the
number of evaluations of  $S(m,x)$ and $\tau(m,x)$
necessary to achieve precise distance inference.

In our  model, the detection threshold $S(m)$ is simply described by a step function while the classifier is treated as a black box for which $\tau(m,x)$ can't be written analytically.
Efficient integration focuses on sampling magnitudes that already satisfy the detection criterion without reference to the classifier, drawn from
\begin{align}  
p(m |  S=1 ,  \mu, p_0) &  = \sum_T p(m,T |  S=1 ,  \mu, p_0)   \nonumber \\
 &  =  \frac{  \sum_T p(S=1 | m,T,    \mu, p_0) p(m | T,  \mu, p_0)  p(T |  \mu, p_0)  }{p(S=1| \mu, p_0)}.
\end{align}
This pdf enters the integrand of $\bar{S}$ (and its derivatives) as
\begin{align}  
p( S=1,{\tau}=1, m, T |  \mu, p_0)  
& = p( {\tau}=1 | S=1,  m, T ,  \mu, p_0)     p(T |  S=1, m ,  \mu, p_0)    \nonumber \\
& \times  p(m |  S=1 ,  \mu, p_0) p(S=1 | \mu, p_0).
\end{align}
 Monte Carlo integration draws
 $N_\text{MC}$ realizations from a fiducial distribution with fixed parameters $\mu_0$, $p_{00}$
 (in the calculations that follow, we  optimistically set $\mu_0$ and $p_{00}$ to their input values),
\begin{equation}
m \sim p(m |  S_{m}=1 ,  \mu_0, p_{00}).
\end{equation}
Then the Monte Carlo integration of $\bar{S}$ is
\begin{align}  
\bar{S}(\mu, p_0)
& \approx \frac{ p(S=1  |  \mu, p_0) }{N_\text{MC}} \sum_{i=1}^{N_\text{MC}}   \sum_{T=0}^1p( {\tau}=1 | S_i=1,  m_i, T ,  \mu, p_0)     p(T | S_i=1, m_i ,  \mu, p_0)  \nonumber \\
& \times  \frac{ p(m_i |  S_{i}=1 ,  \mu, p_{0})}{ p(m_i |  S_{i}=1 ,  \mu_0, p_{00})}.
\end{align}
The partial derivatives of $\bar{S}$ are calculated similarly.

The calculated values of $\bar{S}$ are used to deduce the estimator  and uncertainty of $\mu$.  An independent calculation of $\bar{S}$, using different
realizations for the Monte Carlo integration, produces different values of $\hat{\mu}$ and $\sigma_\mu$.  The standard deviations
of the distributions of  $\hat{\mu}$ and $\sigma_\mu$ from many sets of Monte Carlo realizations represent the
error of interest in this article. 

\section{Examples}
\label{examples:sec}
\subsection{Pure sample}
\label{sec:pure}
Before continuing with the general model, we take a short detour to consider the special case of a Pure sample,
where there are no false positives ($\epsilon_1=0$) and all objects are of type $T=0$.
The classifier nevertheless may not be efficient, $\epsilon_0 \ne 1$, so that the effect of the classifier must  be accounted for.
This example corresponds to an analysis of a spectroscopically pure sample.
This case is useful to review, as the estimator and uncertainty simplify into expressions that may be more easily recognized
by the non-specialist.

There are a few minor changes to the equations shown in the previous section.
\begin{itemize}
\item The likelihood is now predicated on the sample data satisfying both the detection and classification selection criteria
\begin{align}  
L(\mu, p_0; \{m\}) 
& = \bar{S}(\mu, p_0)^{-N}  \prod_{i=1}^N 
R(m_i, \mu, p_0)  
\end{align}
where now
\begin{align}  
\bar{S}(\mu, p_0)
& =\int  p( S=1,{\tau}=1, T=0, m |  \mu, p_0)   dm \nonumber \\
R(m, \mu, p_0)
& =    p(m, S=1,{\tau}=1, T=0|  \mu, p_0).
\label{SRPure:eq}
\end{align}
The change from the general case is that $T$ is no longer a latent parameter, being replaced by $T=0$ and the summation over the two types is removed.
\item Efficient Monte Carlo integration draws from $p(m | S=1 ,   T=0,    \mu, p_0) $,
in terms of which the integrand  of $\bar{S}$ is
\begin{align}  
p( m, S=1,{\tau}=1,T=0 |  \mu, p_0)  
& = p( {\tau}=1 | S=1,  T=0, m,   \mu, p_0)   p(S=1|   T=0,    \mu, p_0)    \nonumber \\
& \times p( T=0|    \mu, p_0) p(m | S=1 ,   T=0,    \mu, p_0) .
\end{align}
\item The pure sample gives no information on the relative rates, so that partial derivative of the likelihood with respect to $p_0$ is  zero.
\item The partial derivatives  of $R$ simplify to
\begin{align}
\frac{\partial \ln{  R}}{\partial \mu} & =  \frac{m-M_0-\mu}{\sigma_0^2} \nonumber \\
\frac{\partial^2 \ln{  R}}{\partial \mu^2} & =  -\frac{1}{\sigma_0^2}.
\end{align}
resulting in an estimator for $\hat{\mu}$ that solves
\begin{align}  
\frac{N}{\bar{S}}  \frac{\partial \bar{S}}{\partial \mu}  \Bigg /  \sum_{i=1}^N  \frac{1}{\sigma_0^2}
+M_0+\hat{\mu}  & =  \sum_{i=1}^N  \frac{m_i}{\sigma_0^2}  \Bigg /  \sum_{i=1}^N  \frac{1}{\sigma_0^2}
\label{estimator_pure:eqn}
\end{align}
and uncertainty estimated from
\begin{align}  
 -\frac{\partial^2 \ln{L}} {\partial \mu \partial \mu} = - \frac{N}{\bar{S}^2} \left( \frac{\partial \bar{S}}{\partial \mu}\right)^2    + \sum_{i=1}^N  \frac{1}{\sigma_0^2}
 \label{hessian_pure:eqn}
 \end{align}
 evaluated at $\hat{\mu}$.
If the sampler were complete, then $\partial \bar{S}/\partial{\mu}=0$ and the estimator in Eq.~\ref{estimator_pure:eqn} would simplify to being the weighted
mean of the measurements with statistical uncertainty $\sigma_0/\sqrt{N} $ from Eq.~\ref{hessian_pure:eqn}. 
 \end{itemize}

The estimator $\hat{\mu}$ and its statistical uncertainty $\sigma_{\mu}$ are calculated from Eqs.~\ref{estimator_pure:eqn} and \ref{hessian_pure:eqn}
respectively for many instantiations of Monte Carlo draws.
The standard deviations in these calculated values  represent
their errors 
due to MC integration.

To illustrate, we take an example setting the absolute magnitude to $M_0=0$, the magnitude dispersion to $\sigma_0=0.1$~mag,
the classifier efficiency to $\epsilon_0=0.95$, and no false positives $\epsilon_1=0$.  The input model parameters for the distance modulus
and Population-0 fraction
are $\mu=0$ and $p_0=0.5$ respectively.

Both $\hat{\mu}$ and $\sigma_\mu$ depend on the sample data through $R$. 
In place of instantiating a simulated data set for these calculations, we adopt
a ``typical'' dataset that satisfies
 \begin{align}
 \sum_{i=1}^N  \frac{m_i}{\sigma_0^2}=
\bigg \langle \sum_{i=1}^N  \frac{m_i}{\sigma_0^2} \bigg \rangle & =-  \frac{\mathcal{N}(m_\text{lim} -M_0-\mu_0, \sigma_0)} {\text{cdf}(m_\text{lim} - M_0-\mu_0, \sigma_0 )},
 \end{align}
where ``$\text{cdf}$'' is the cumulative distribution function of a Normal distribution.  (Recall that the examples adopt classification probabilities uncorrelated with $m$.)
This choice eliminates bias in the best-fit parameter values due to the statistical noise in a single realization of data, allowing us to  focus  on the errors due to MC 
integration.

\begin{figure}
\centering
\includegraphics[width=11cm]{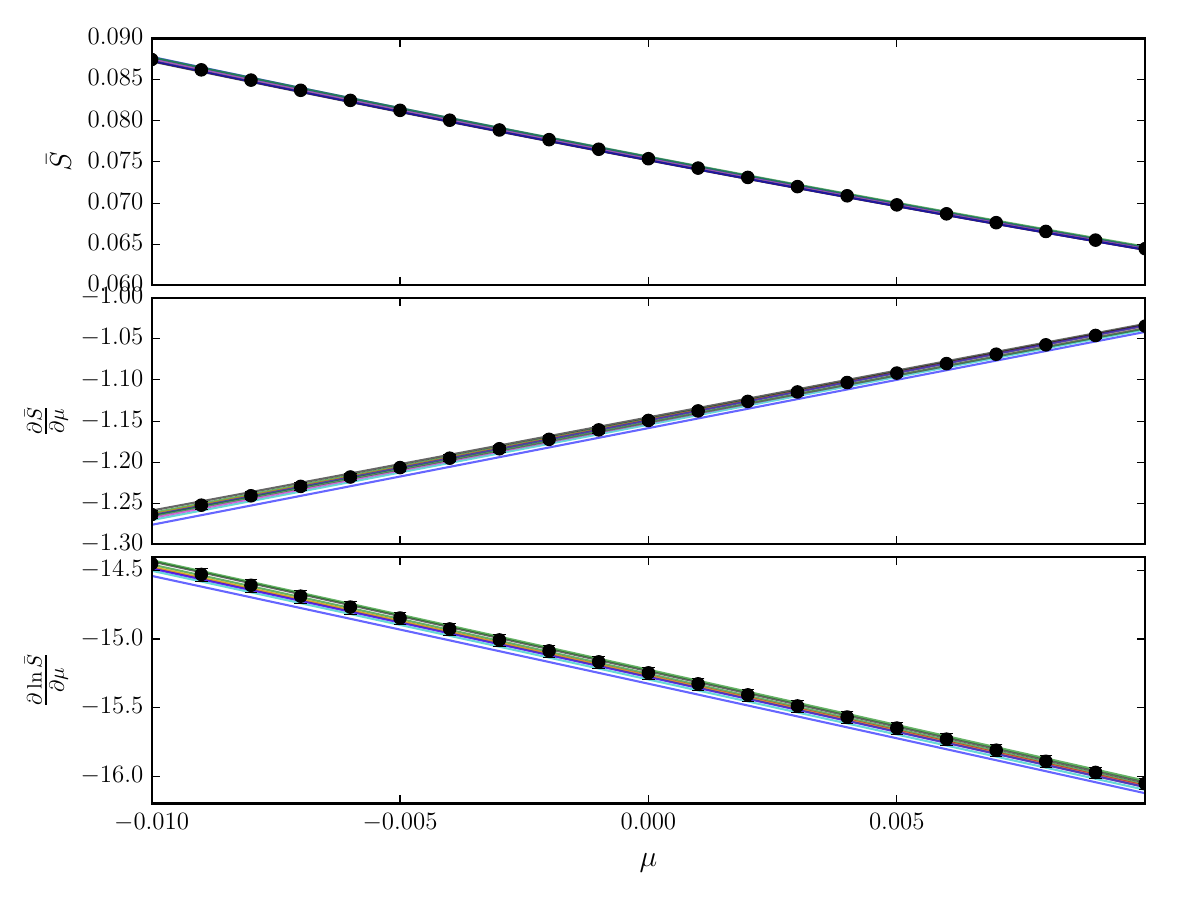}
\caption{Points with error bars are the 16, 50, and 84\%-iles  of the calculated values of $\bar{S}$, $\partial \bar{S}/\partial \mu$, and $\partial \ln{\bar{S}}/\partial \mu$
for $m_\text{lim}=-0.1$~mag and 
independent draws of $N_\text{MC}=10,000$.  Several instantiations of these functions are overplotted as solid lines.
\label{barSnorm:fig}}
\end{figure}

The errors due to Monte Carlo integration in
the calculation of  $\bar{S}$ and its partials are shown in Figure~\ref{barSnorm:fig}, which plots the 16, 50, and 84\%-iles
of the calculated values $\bar{S}$, $\partial \bar{S}/\partial \mu$, and $\partial \ln{\bar{S}}/\partial \mu$ 
for $m_\text{lim}=-0.1$~mag and 
independent draws of $N_\text{MC}=$10,000.  Several instantiations of these functions are overplotted, showing that the errors in the function values
at different $\mu$ are correlated.   The curves are linear given the choice of a classifier that is uncorrelated with magnitude; classifier output that
are correlated with magnitude would imprint $\mu$-dependent structure in these curves.

\begin{figure}
\centering
\includegraphics[width=18cm]{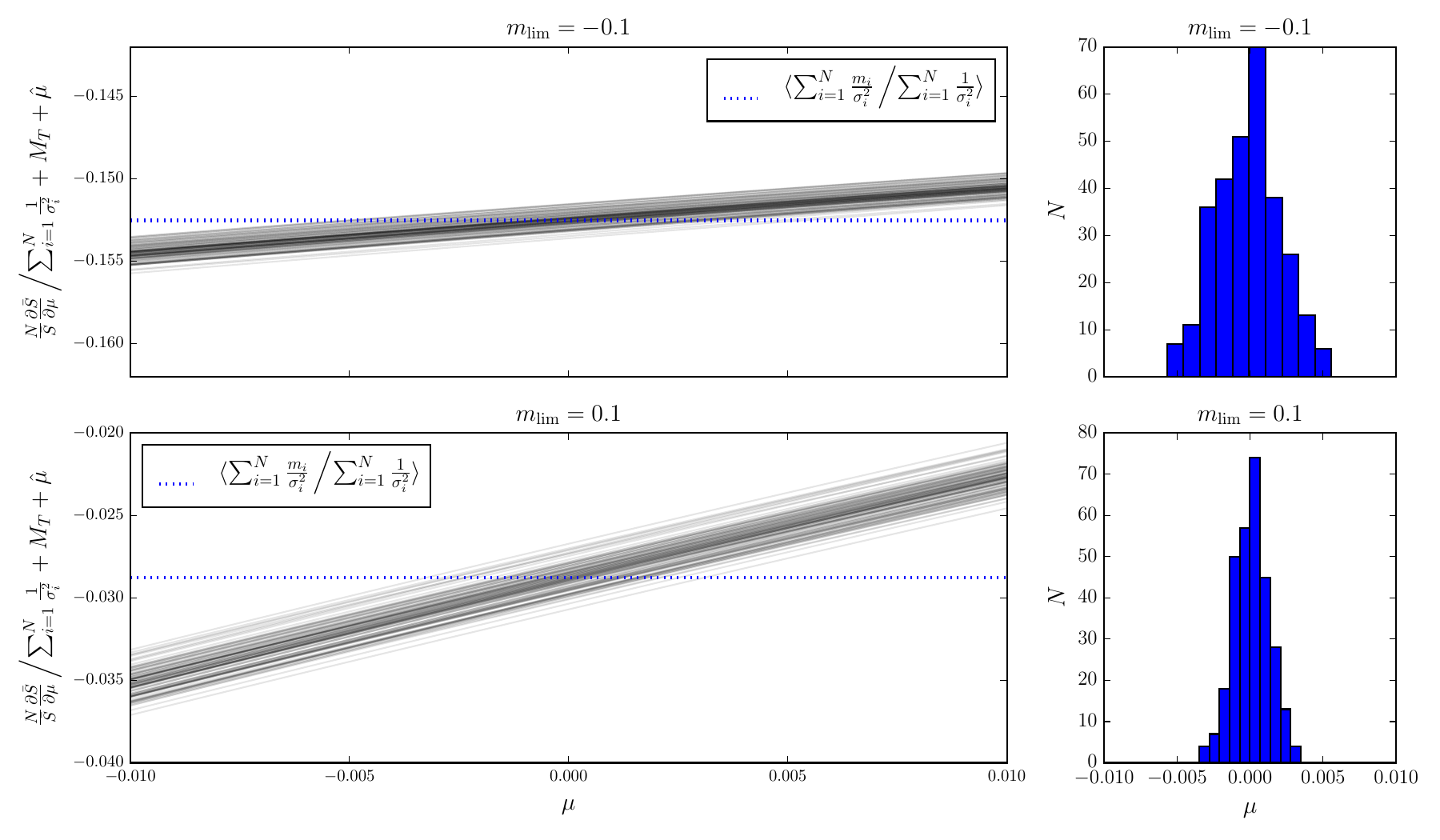}
\caption{Left:  For the pure sample example, the left-hand term of Eq.~\ref{estimator_pure:eqn}, $\frac{N}{\bar{S}}  \frac{\partial \bar{S}}{\partial \mu}  \bigg /  \sum_{i=1}^N  \frac{1}{\sigma_0^2}
+M_T+\hat{\mu}$, for  a number of independent MC realizations. 
The line thickness for each realization is the same, the apparent non-uniformity is due to the density of the lines.
The dotted horizontal lines
are the right-hand terms, $ \sum_{i=1}^N  \frac{m_i}{\sigma_0^2}  \bigg /  \sum_{i=1}^N  \frac{1}{\sigma_0^2}$, for an ``average'' dataset.
The estimator $\hat{\mu}$ for a single MC is where the curve and the horizontal line intersect. 
Right: Histograms of the estimators of $\hat{\mu}$ (where the sloped curves intersect the horizontal line in the left-hand plots) for a number of independent MC realizations.
\label{estimatornorm:fig}}
\end{figure}

The estimator $\hat{\mu}$ is  the value of  $\mu$ where the term on the left  of Eq.~\ref{estimator_pure:eqn}
is equal to the data-dependent, $\mu$-independent term
on the right.  Figure~\ref{estimatornorm:fig} plots the left-hand term for  a number of instantiations of the MC integration, and
the the constant values of the right-hand terms as
 horizontal lines.  The estimators, where the sloped curves intersect the horizontal line, are different for each MC
realization, and the estimator errors are given by the standard deviations of these intercepts.

We call attention to several interesting features seen in Figure~\ref{estimatornorm:fig}.
The intercepts are centered around  the input $\mu=0$, confirming that the $\partial \ln{\bar{S}}/\partial{\mu}$ term in Eq.~\ref{estimator_pure:eqn}  properly accounts for
the Malmquist bias in the magnitude-limited sample.
Within  the range of
the plot, the sloped curves are nearly parallel, meaning that the Monte Carlo errors are not so sensitive to the data, i.e.\ a different value of the right-hand term and location
of the horizontal line.  The slope of the left-hand term varies with the magnitude limit $m_\text{lim}$, from 1 when $\partial{\bar{S}}/\partial \mu =0$, decreasing to 0 as the
 threshold gets more severe.  Shallower slopes  are responsible for larger errors in the intersection of the horizontal line and hence in  $\hat{\mu}$.
We again emphasize that lack of structure in the left-hand term is a feature of the classifier considered.  A classifier that has
$\partial \ln{\bar{S}}/\partial \mu + \mu$ that is not monotonic in $\mu$ can have a likelihood with multiple local maxima whose characterization
requires more detailed analysis than used here.

\begin{figure}
\centering
\includegraphics[width=11cm]{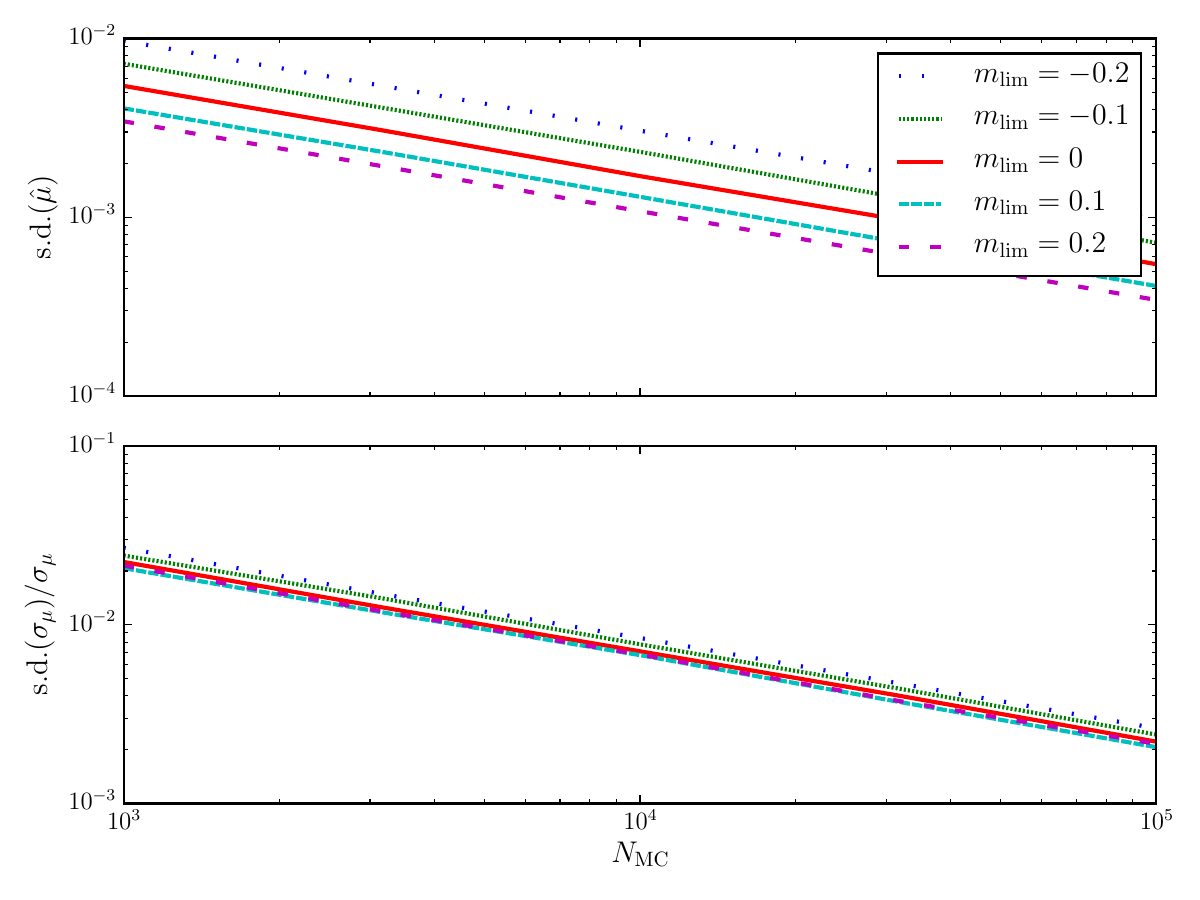}
\caption{Errors in the estimators, $s.d.(\hat{\mu})$, and the fractional errors in the $\mu$ uncertainty,  $s.d.(\sigma_{\mu})/\sigma_{\mu}$,
 as a function of $N_\text{MC}$ for the Pure sample example.
\label{sigmasPure:fig}}
\end{figure}

The errors in the estimators, written as $s.d.(\hat{\mu})$ as a function of $N_\text{MC}$  are calculated for several scenarios.
The fractional errors in the $\mu$ uncertainty derived from the Hessian, written as $s.d.(\sigma_{\mu})/\sigma_{\mu}$, are also calculated.
Results for a range of $m_\text{lim}$ are shown  in Figure~\ref{sigmasPure:fig}.
For a given $N_\text{MC}$,  errors are proportional to  $N_\text{MC}^{-1/2}$ as expected from Monte Carlo integration.

As a function  of limiting magnitude, the more complete the sample (large limiting magnitude), the smaller the error in the estimators.
As the sample reaches completeness, the $\bar{S}$ terms and their errors become relatively unimportant.
Recall that the relevant statistic is  $ \frac{1}{\bar{S}} \frac{\partial \bar{S}}{\partial \theta}$ whose uncertainty
is the integration error in $\frac{\partial \bar{S}}{\partial \theta}$ further   weighted by $\bar{S}^{-1}$. 
The Monte Carlo integration draws from $p(m | S=1 ,    \mu, p_0)$ with variance $\int^{m_\text{lim}}_{-\infty} (m^2-\bar{m}) \mathcal{N}(m-M_0-\mu_0,\sigma_0) dm$ that steadily drops with decreasing (more severe) $m_\text{lim}$.  At the same time
$\bar{S}= \int^{m_\text{lim}}_{-\infty} \mathcal{N}(m-M_0-\mu_0,\sigma_0) dm$ also  drops with decreasing $m_\text{lim}$.
Although the narrower magnitude dispersion of a truncated dispersion gives lower integration errors for $\partial{\bar{S}}/\partial{\mu}$,
the weighting by $\bar{S}$ in fact leads to the larger errors  seen in Fig.~\ref{estimatornorm:fig}.)

While the same trend is true for the error in the uncertainty $s.d.(\sigma_{\mu})$,  $s.d.(\sigma_{\mu})/\sigma_{\mu}$
does not monotonically decrease with  increasing limiting magnitude  because $\sigma_\mu$  decreases even faster 
with limiting magnitude.

For a survey with a target precision in $\mu$, the required number of Monte Carlo samples $N_\text{MC}$ can be read from
Figure~\ref{sigmasPure:fig}.  For example, the uncertainty is $<0.01$~mag for $N_\text{MC}= 1,000$ with limiting magnitude
as bright as $m_\text{lim}=-0.2$~mag.  Correct calculation of parameter uncertainties is as important as the estimators themselves.
A 1\% error fractional uncertainty is obtained with $N_\text{MC}= $10,000.

\subsection{General example}
We now return to the general case of a Two-Population model with sample contamination.
The true type of each object is uncertain, so the latent parameter $T$ comes into play.
The data can longer be isolated into a parameter-independent term as was the case for  Pure sample.  The relative rate of the two populations $p_0$ must now be considered.
This example corresponds to an analysis that would be undertaken when the sample is selected using  an imperfect photometric classifier.

The example considered here is similar to that of the Pure sample.   Population 0 
has absolute magnitude $M_0=0$ and magnitude dispersion of $\sigma_0=0.1$~mag.
The contaminating population has $M_1=1$ and magnitude dispersion of $\sigma_1=0.5$~mag.
The classifier efficiency remains $\epsilon_0=0.95$, but now the outlier population can pass the classification
criteria with probability $\epsilon_1=0.05$.  The fiducial model parameters
are $\mu=0$ and $p_0=0.5$.

Results are presented for one realization of the data.  They yield a best-fit values that are 
not precisely the fiducial parameter values but are within statistical uncertainties.
The sample data are selected from 10,000 objects from both populations that satisfy the magnitude-detection threshold,
which are subsequently filtered by the classifier.

\begin{figure}
\centering
\includegraphics[width=18cm]{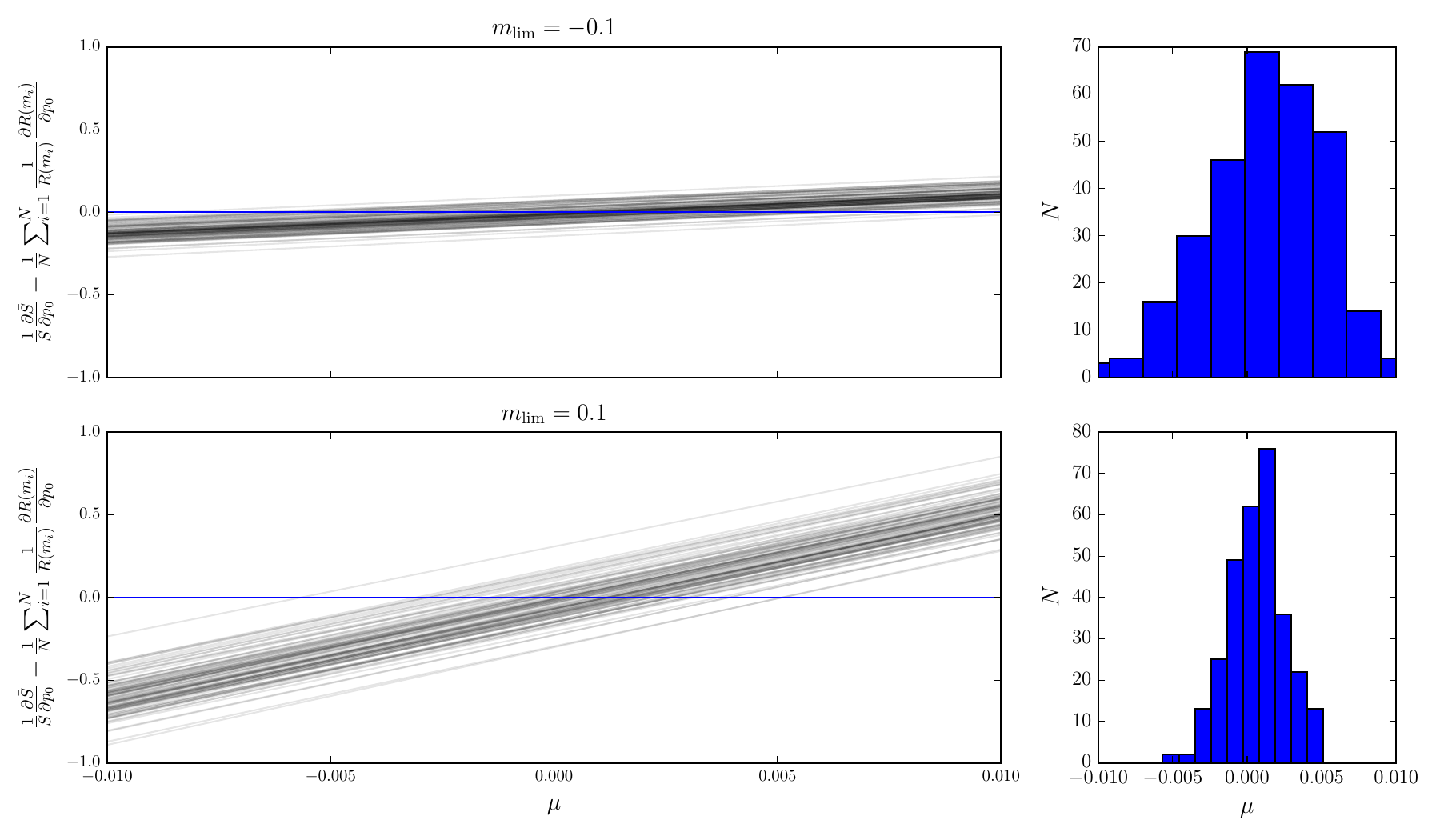}
\caption{Left: For the general example, the partial derivative of the log-likelihood $\frac{1}{\bar{S}} \frac{\partial \bar{S}}{\partial \mu}
- \frac{1}{N} \sum_{i=1}^N \frac{1}{R(m_i)} \frac{\partial R(m_i)}{\partial \mu}$, whose root is the estimator $\hat{\mu}$, plotted for a number of realizations
of the Monte Carlo integration with $N_\text{MC}=10,000$.  The dotted horizontal lines are at zero, and are shown to help indicate the locations 
of the roots from the different realizations.
Right: Histograms of the estimators of $\hat{\mu}$ (where the sloped curves intersect the horizontal line in the left-hand plots) for a number of independent MC realizations.
\label{estimatorselection1:fig}}
\end{figure}

\begin{figure}
\centering
\includegraphics[width=18cm]{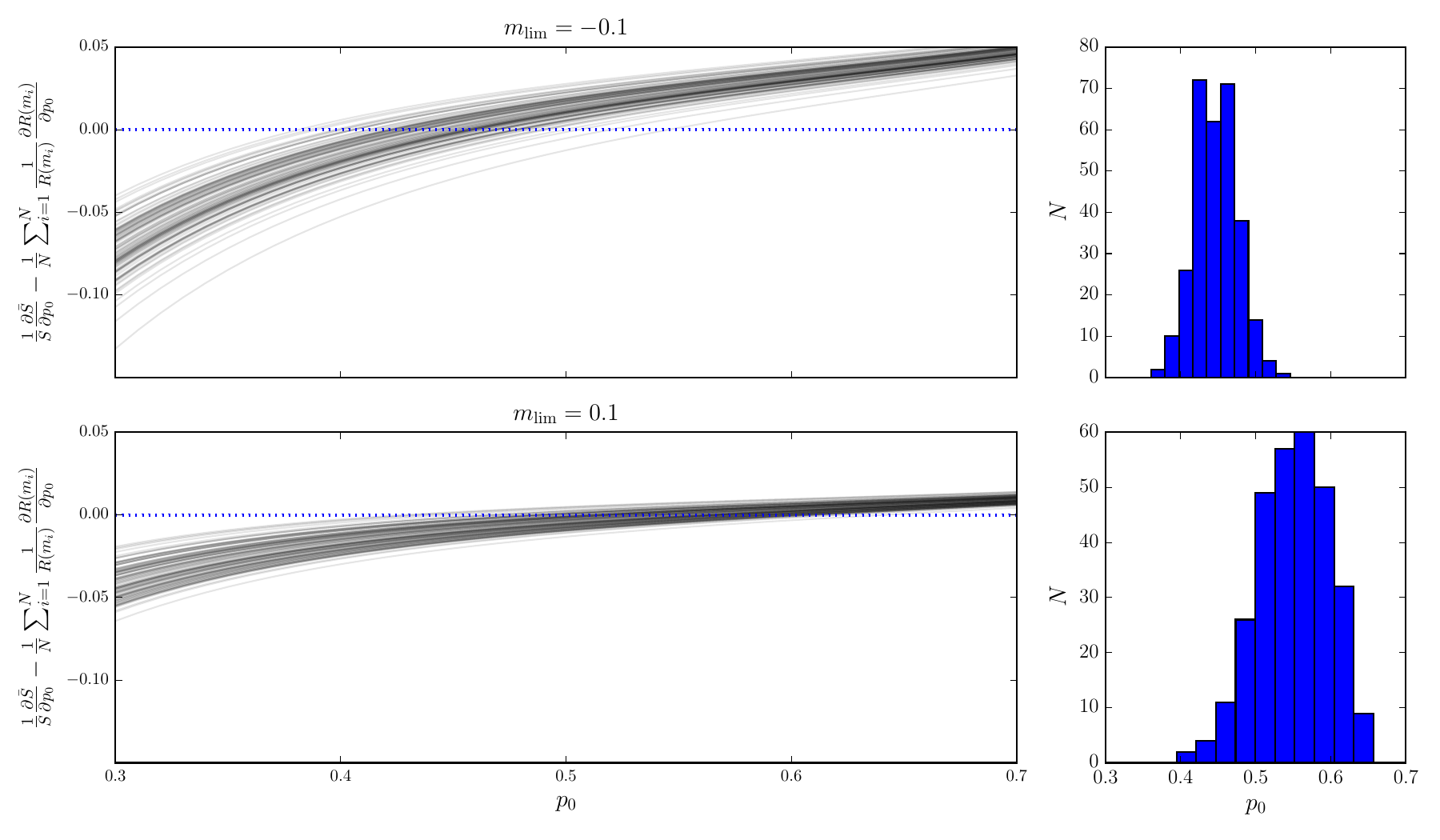}
\caption{Left: The partial derivative of the log-likelihood $\frac{1}{\bar{S}} \frac{\partial \bar{S}}{\partial p_0}
- \frac{1}{N} \sum_{i=1}^N \frac{1}{R(m_i)} \frac{\partial R(m_i)}{\partial p_0}$, whose root is the estimator $\hat{p}_0$, plotted for a number of realizations
of the Monte Carlo integration with $N_\text{MC}=10,000$.  The dotted horizontal lines are at zero, and are shown to help indicate the locations 
of the roots from the different realizations.
Right: Histograms of the estimators of $\hat{p}_0$ (where the sloped curves intersect the horizontal line in the left-hand plots) for a number of independent MC realizations.
\label{estimatorselection2:fig}}
\end{figure}

The estimator  $\hat{\mu}$ solves Eq.~\ref{estimator:eqn}, and is calculated
for many
different instantiations of  Monte Carlo integrals  with $N_\text{MC}=$10,000
and plotted in 
Figure~\ref{estimatorselection1:fig}.  The average solution is centered near the fiducial $\mu=0$.
The population-ratio parameter $p_0$ has to be accounted for
and the distribution of its estimator $\hat{p}_0$ for the  instantiations of  Monte Carlo integrals is
plotted in 
Figure~\ref{estimatorselection2:fig}.
The average solution is centered near the fiducial $p_0$.  
(Recall that results depend on the draw of the data, which induces a small bias relative to the input parameters.)
The dispersion in the $\hat{p}_0$ is relatively large; in fact for $N_\text{MC}=$1,000, 
a significant fraction of the calculated partial derivatives of the log-likelihood do not have a root within the allowed range $[0,1]$.
The error in $\hat{p}_0$ is smaller for $m_\text{lim}=-0.1$ than it is for $0.1$, a result of having a larger fraction of 
Population 1 objects that pass the magnitude cut (the broad bright tail of Population 1 contributes relatively more than 
the narrow tail of Population 0), which leak into the classified sample.  

\begin{figure}
\centering
\includegraphics[width=11cm]{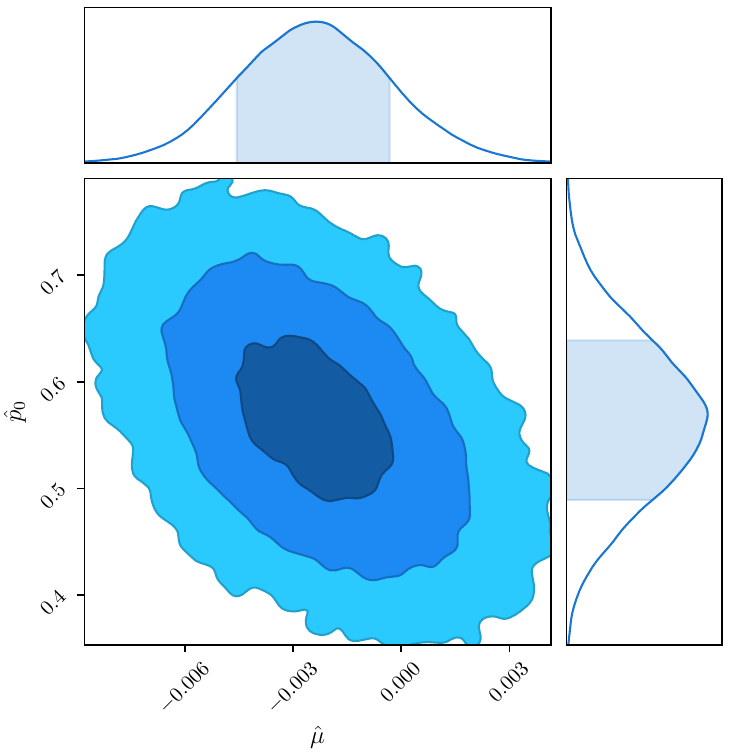}
\caption{Distribution of the estimators $\hat{\mu}$ and $\hat{p}_0$ from many
different instantiations of the
 Monte Carlo integrals with $N_\text{MC}=10,000$.  Shaded regions in the histograms correspond to 1$\sigma$, those in the contour plot to 1, 2, and 3$\sigma$ in one dimension.
\label{surface:fig}}
\end{figure}

The estimators $\hat{\mu}$ and $\hat{p}_0$ are calculated for 
many instantiations of the
 Monte Carlo integrals and their distribution is shown in Figure~\ref{surface:fig}.
The contour is not centered at the input value due to the statistical noise inherent in the data realization for which the calculation is performed.
The fit parameters are correlated, as seen in the rotated  shape of the ellipses.   

\begin{figure}
\centering
\includegraphics[width=11cm]{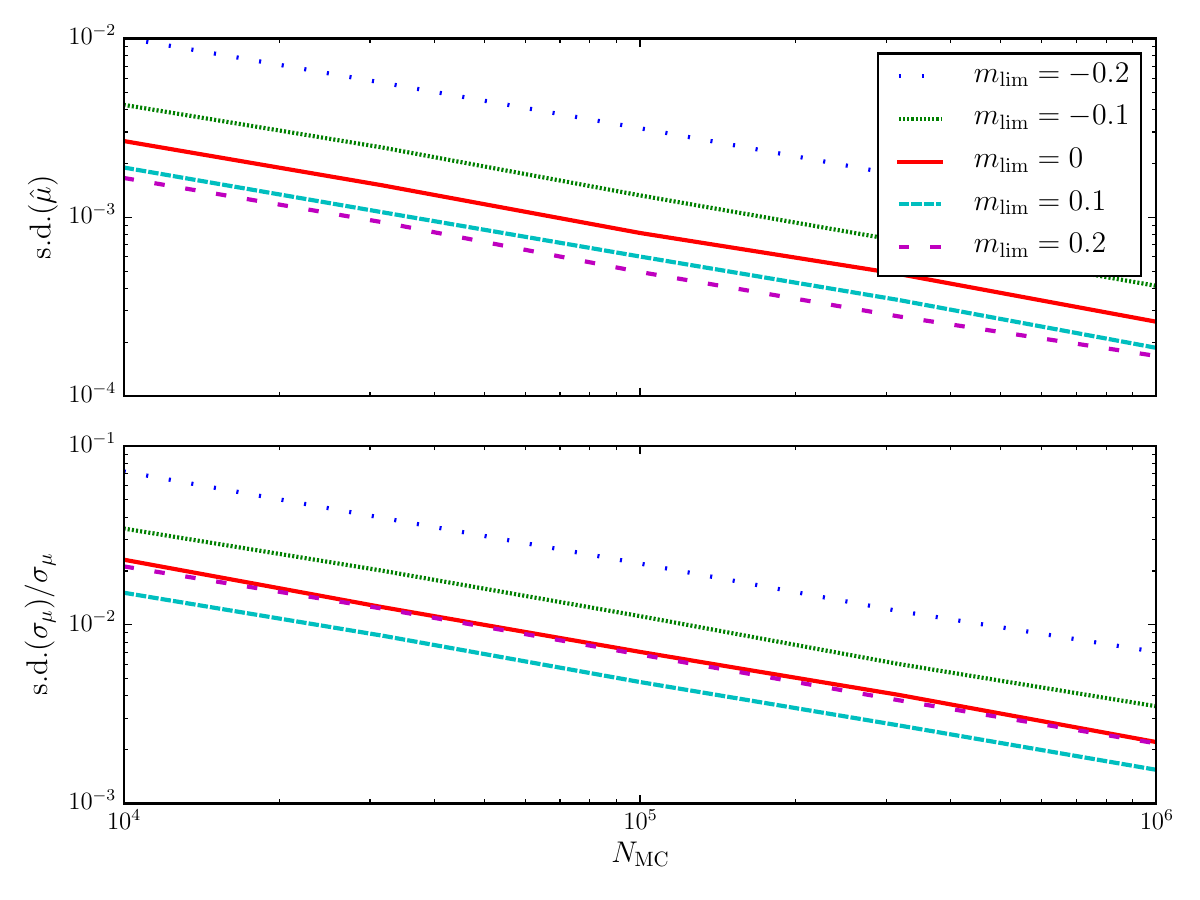}
\caption{Errors in the estimators, $s.d.(\hat{\mu})$, and the fractional errors in the $\mu$ uncertainty,  $s.d.(\sigma_{\mu})/\sigma_{\mu}$,
 as a function of $N_\text{MC}$ for the Two-Population example.
\label{sigmas:fig}}
\end{figure}

The errors in the estimators and their uncertainties, $s.d.(\hat{\mu})$ and $s.d.(\sigma_{\mu})/\sigma_{\mu}$, are calculated for a range of scenarios as a function of $N_\text{MC}$
and presented
 in Figure~\ref{sigmas:fig}. 
Qualitatively the results are similar to those for the Pure sample in \S\ref{sec:pure}.  They span similar ranges of uncertainty, and have the
same dependency on $N_\text{MC}$.  
Both examples share the same sequence of $m_\text{lim}$ when ordering error sizes, though they do differ somewhat in their quantitative 
details.

In this case with two populations with different magnitude distributions and different probabilities of being classified as Population 1,
the fraction of Population 1 and 2 objects entering
the sample changes as a function of $m_\text{lim}$.

Comparison of the Two-Population scenario of this section and the pure sample in \S\ref{sec:pure} shows that
the former requires an order-of-magnitude larger $N_\text{MC}$ in order to achieve the same precisions.
The $m_\text{lim}$-dependence of the  variance of the MC sampling distribution  $p(m | S=1 ,    \mu, p_0)$  is quite different
than before,
decreasing as the magnitude limit is turned on, reaching a minimum, and then increasing with increasing brightness cutoff
when the sample predominantly consists of the bright tail of the second population.
At $m_\text{lim}=0$~mag, the variance of the  two-population distribution is 1.4$\times$ larger than that of the pure
distribution.  Given that a significant fraction of the faint population does not pass sample selection,
 the $\bar{S}$  is $\approx 0.475$ that of the pure distribution.  Together these factors would account for a
 $6.2\times$ larger $N_\text{MC}$  for the two-population model to match the parameter precision of a pure model.
 Of course a complete accounting of the comparison of the $N_\text{MC}$ of the two models depends on the integrands
 that appear in the partial derivatives of $\bar{S}$.

\section{Discussion}
\label{discussion:sec}

Model inference and parameter constraints require the evaluation of an integral, which generally is not analytic and so is solved numerically.
The integrand contains the terms $S(m)$ and $\tau(m,x)$, which represent sample and classification selections respectively.  At the time
of analysis (say after the conclusion of the LSST survey), these functions whose properties will have been set long previous
(say Year 1 of LSST), need to be evaluated many times.
Understanding the computational implications for this aspect of the analysis and consequently the requirements for Brokers  was the prime motivation of this work.

Conservatively speaking,
both software and data states that influence sample and classification selections non-trivially, be they from
the Project, third-party Brokers, or in-house pipelines, should be available for execution $\sim 10$ years after real-time usage to feed cosmology analysis.
We thus advocate the  packaging of
software in containers and the tracking data state histories, so that the processing at any given moment can be rerun at a later date on  a different computer.

In the examples fitting for a single $\mu$ considered in \S\ref{examples:sec}, the number of function evaluations needed to get per-mil precisions is on the order of millions.  
Expanding the analysis by splitting the sample into ten independent redshift bins gets the number of evaluations up to the ten million real alerts per night expected from LSST.
Nevertheless, the rich dataset from LSST can lead to more ambitious analysis and complex models to accommodate considerations such as photometric redshifts,
non-Gaussian magnitude distributions of  backgrounds that mimic SNe~Ia, SN~Ia subtypes.

\section{Conclusions}
\label{conclusions:sec}
The likelihood of an experiment that has sample selection includes an integral over all possible objects that could have but did not enter
the sample. 
Monte Carlo integration is a viable way to
 numerically evaluate that integral.
Being in the integrand, the sample-selection and classification efficiencies are  represented by their values at the random points generated by the Monte Carlo.
 The precision of the integration
 depends on the number of points at which the integrand, and hence the efficiency, is sampled.
Integration errors propagate into errors in the parameter estimators
and the fractional errors in the parameter uncertainties; these uncertainties are independent of the number of objects in the analysis sample.
 Requirements of parameter
 estimation thus set the minimum number of points at which the efficiency must be evaluated.

For several example scenarios for measuring distances with SNe~Ia, 
the sample selection efficiency must be evaluated for a number of simulated supernovae that far
exceeds that of the sample itself.
Otherwise the uncertainty in the distance estimator $\hat{\mu}$ due to the miscalculated likelihood can 
be a significant fraction of the overall $\mu$ uncertainty. 

The  number of Monte Carlo evaluations needed to meet precision requirements is sensitive to the model.  
The numerical evaluation of the integral needed to calculate the optimal estimator (e.g., Eq.~\ref{estimator:eqn})
has variance $ \propto \frac{\sigma^2_N}{N}$, where $\sigma^2_N$ is the variance of the drawn values of 
the integrand.
When the underlying model includes multiple source populations, the pdf of the observables $p_y$ 
broadens, the pdf of the draws of the integrand broadens, and the variance of $\bar{S}$ gets larger, leading
to an increase in the error of $\mu$ found in the root solution.

A survey's statistical performance improves when an informative classifier is used; in this way classifiers indirectly affect
the required number of evaluations of the selection function.  Characterizing the classifier itself requires
simulated SNe, though in the formalism we present here their sample selection need not be assessed.  Determining the  required precision
on classifier performance
is beyond the scope of this study.

{\it Further Work}: More work is necessary to create a realistic LSST SN analysis
needed to produce sample-selection requirements that will be used for classifier evaluations.
The toy models presented in this note are extreme simplifications of the one that will eventually be used to describe the LSST sample.
Here we treated SNe~Ia as standard candles but they are in fact standardizable candles with their own internal subparametrization.  While we used a classification selection that is
$m$-independent, the first and second moments of the sample and underlying population magnitudes will probably differ and induce biased sample magnitudes.
It is important to consider selection bias as it enters through redshift measurements: typically redshifts are obtained from the host galaxy, but when the host is too faint
a redshift may be inferred from the supernova itself or the supernova may be excluded from the sample.  Given that
SN models correlate intrinsic supernova and
host-galaxy properties, proper accounting for redshift sample selection should be implemented in these methods.
The sample supernovae were taken to have the same
observables whereas it is expected that there will be a subset of spectroscopically typed objects (with its own sample selection function) analyzed together
with the photometric set.
The magnitude distribution of supernovae was taken to be Normal, whereas the parameter-distributions of SNe~Ia are not Normal and indeed
must be modeled in the fit.
The observed SN~Ia magnitude dispersion (before color and light-curve shape corrections) relevant for a magnitude-limited survey is $\sim 0.4$~mag,
larger than the 0.1~mag considered in our model that omits the details of SN magnitude corrections. 
An implication is that SN~Ia subpopulations must also be included in the model.
The true sample selection will be date-dependent, depending on the conditions of actual observing, and may be stochastic (i.e.\ $0<S(m)<1$).
Multiple sample selections may enter a single analysis, for example a training set of follow-up supernovae can be identified using only
early data whereas the larger sample of supernovae would use data covering the full light-curve evolution.
Here we considered the distance $\mu$ as the parameter of interest, whereas 
many experiments are evaluated based on 
the physically relevant dark energy equation of state parameters
$w_0$ and $w_a$. 
The more diverse the model for the underlying supernova populations,
the more  possible realizations there are  to integrate over, which should increase the Monte Carlo sampling requirements
to maintain precision in the likelihood. 
As the community continues to develop future analyses, the implications for characterizing the sample selection should be
kept in mind.

\section*{Acknowledgements}
This paper has undergone internal review in the LSST Dark Energy Science Collaboration.
The internal reviewers were Rahul Biswas, Phil Marshall, and Gautham Narayan.
This work is supported by 
the U.S.\ Department of Energy, Office of Science, Office of High Energy 
Physics, under contract No.\ DE-AC02-05CH11231. 
The DESC acknowledges ongoing support from the Institut National de Physique Nucl\'eaire et de Physique des Particules in France; the Science \& Technology Facilities Council in the United Kingdom; and the Department of Energy, the National Science Foundation, and the LSST Corporation in the United States.  DESC uses resources of the IN2P3 Computing Center (CC-IN2P3--Lyon/Villeurbanne - France) funded by the Centre National de la Recherche Scientifique; the National Energy Research Scientific Computing Center, a DOE Office of Science User Facility supported by the Office of Science of the U.S.\ Department of Energy under Contract No.\ DE-AC02-05CH11231; STFC DiRAC HPC Facilities, funded by UK BIS National E-infrastructure capital grants; and the UK particle physics grid, supported by the GridPP Collaboration.  This work was performed in part under DOE Contract DE-AC02-76SF00515.


\end{document}